\RequirePackage{amsmath}
\documentclass[a4paper]{llncs}
\usepackage{cite}
\usepackage{amsmath,amssymb,amsfonts}
\usepackage{algorithmic}
\usepackage{graphicx}
	\graphicspath{ {Images/} }
\usepackage{textcomp}
\usepackage{xcolor}
\usepackage[english]{babel}
\usepackage{xspace}
\usepackage{lmodern}            		
\usepackage[pdfencoding=auto]{hyperref}	
\usepackage{url}
\usepackage{color}
\usepackage{enumitem}
\usepackage{adjustbox}
\usepackage{array}
\usepackage{tabularx}
\usepackage{makecell}

\pdfoutput=1

\newcolumntype{R}[2]{
    >{\adjustbox{angle=#1,lap=\width-(#2)}\bgroup}%
    l%
    <{\egroup}
}
\newcommand*\rot{\multicolumn{1}{R{45}{1em}}}

\def\BibTeX{{\rm B\kern-.05em{\sc i\kern-.025em b}\kern-.08em
    T\kern-.1667em\lower.7ex\hbox{E}\kern-.125emX}}

\begin{document}

\title{Modeling and verification method for an \\early validation of a train system}

\author{Ronan Baduel\inst{1,2} \and
Iulian Ober \inst{2} \and
Jean-Michel Bruel \inst{2}}

\authorrunning{R. Baduel et al.}

\institute{Bombardier Transport, 59154 Crespin, France \and University of Toulouse - IRIT, 31400 Toulouse, France, email: \email{name.surname@irit.fr}
}

\maketitle

\begin{abstract}
This paper presents the results achieved while pursuing the verification and validation of a train system behavior at the first steps of development in an industrial context. A method is proposed, supported by preliminary results through the definition and verification of constrained states and preconditions to use cases, as well as a structure for the behavior. 
\end{abstract}

\section{Introduction}

How to perform validation on a system behavior is a field of research in systems engineering \cite{fujimoto_research_2017}. Several techniques exist, such as tests, simulation, model checking, etc. One key aspect, however, is that these techniques tend to be deployed on design or implementation of a system, in order to check an \emph{integrated} behavior that may not have been fully specified in the first place. A general truth in system V\&V is that the earlier an error is detected, the lower will be the cost \cite{stecklein_error_2004}. The most critical errors are those made when expressing the requirements. Works \cite{abrial_modeling_2010, thiago_rocha_silva_2018, benveniste_contracts_2015} show that one should specify, and if possible validate, the \emph{expected} behavior of the system \emph{as a whole} using requirements and scenarios, before any design or implementation. The main issue encountered to achieve such a task is that a model of the system is required to define and support the behavior.

In this paper, we present early validation results obtained thanks to the definition of global states and modes describing a train and its behavior at operational level. This work is conducted in the scope of a project in Bombardier Transport (BT), a train manufacturing company. This project aims at establishing a continuous validation method along a train system development process. The target of the validation is the system behavior, which we define as the way the system reacts under given circumstances.

\subsection{Context}

Over the years, BT has developed its own modeling method and SysML profile to develop train system behaviors \cite{Chami_etal_BT2015}, based on existing approaches \cite{lamm_functional_2010}. The first step in the specification, once the requirements have been analyzed, is to define scenarios and the train life-cycle highlighting the services provided by the system, and in which circumstances. The system behavior is to be validated. It requires having an executable model of the whole system, and is currently done later in the development process, using implemented software and simulated hardware. 

The development process and the V\&V activities at system level are separated in two BT teams: one specifying the system, and one conducting V\&V activities through co-simulation. The first team has skills in requirements analysis, functional specification and SysML modeling. The second focuses on tests, programming and simulation. Those two teams have separate responsibilities, if only due to the organization and according to good practices.

The behavior of a train system is defined through hundreds of use cases, classified among hundreds of \emph{scopes}. A \emph{scope} is part of a functional breakdown structure that classifies the use cases and the functions according to their domain (e.g. energy, traction, etc.). The scopes are divided among different requirements and functional engineers. In the chosen metro \emph{MOVIA} case study, the specification at operational level includes 277 use cases contained in 60 root scopes of a classification system, divided among 12 functional and requirements engineers. 

Each use case is described by a sequence diagram. Redundancy in specification is avoided by having each engineer work on dedicated scopes. The inconvenient is that they define behaviors separately, using a non-formalized nor centralized knowledge regarding the system. Consequently, the specifications are unrelated, without any integration. Rather than just specifying what the system-to-be does, we have to specify ``what'' system we want to obtain \cite{soeken_formal_2015}, meaning an abstract model of the system induced by the specifications. Such a model implement an integrated behavior by conditioning the use cases and track their effect on the evolution of the system state, without specifying how those use case are realized inside the system.

Regarding the V\&V activities, there are limitations in BT. Executable models such as grafcet \cite{baracos_grafcet_1992} have been used where parts were missing in the co-simulation, but they correspond to designs provided by subsystems developers or external providers. There is currently no solution nor models to check the system before any implementation. 

It is currently possible to check that the system does what it is expected to do through tests, later in the process, using co-simulations or bench tests. However, there are no practical solution for capturing unknown or unwanted behavior. While it is possible to generate random inputs in a co-simulation, all cases disproving a property have to be analyzed by an engineer to assess its relevance. BT stopped using such a solution, as experience showed that checking a property could result in hundreds or thousands of cases to analyze, most of them irrelevant as they suppose a use of the train system that cannot happen in reality. There is a need to constrain either the system behavior or its inputs.

\subsection{Issues}

Each functional engineer has a knowledge on a specific scope of the train, and uses informal information from textual requirements and her own experience to make specifications. There is no integration by the models. The overview of the whole system is done informally by an individual. This leads to a lack of a formal description of the system at specification level, which prevents its automatic verification \cite{soeken_formal_2015}. There is a need to express and check a common information regarding the whole system. The solution should verify automatically the coherence of the information, so that engineers do no depend on the validation team for their specifications. On the other hand, the validation team should receive an integrated, formalized and verified information to build an executable model, rather than interpreting it on their own. 

The integration of properties and behaviors results in a phenomenon called \emph{emergence} \cite{abbott_emergence_2007}, as the system is more than the sum of its elements. There are potential unknown and unwanted aspects of the system. Integrating the system implies specifying the integration to avoid emergence. Specifying the integration is the role of the functional engineers, not of the validation team, which is currently the one doing the integration when asked for a model of the system.

The different issues can be listed as a lack of:

\begin{itemize}
\item formalized requirements and properties for validation purposes,
\item specification on the preconditions and circumstances of execution of system functions,
\item integrated representation of the system,
\item solutions to structure the behavior before making a design,
\item solutions to checking possible (unwanted) behaviors.
\end{itemize}

The goal of the proposed approach is not to give an optimized solution to the V\&V of the specification at a given level of development, or even globally, but to provide a way to continuously conduct these activities along the development process and with traceability of both V\&V requirements, results and models, based on an existing modeling method. Accessibility, simplicity and quickness is preferred over exhaustivity and formalism, taking into account the industry needs and capabilities. A given solution cannot be immediately implemented across the entire development process. It has to be gradual, following the evolution of both the modeling method and the development process.

\subsection{Related works}

The proposed description of the system and its behavior is done using SysML through state machines. The notion of state can be traced to the general system theory \cite{le_moigne_theorie_1994}. Is is explained that defining the state of a system at a point in time is necessary to express its behavior, which is linked to its evolution. State machines have long been used to define and check system behaviors \cite{harel_statecharts_1987, hopcroft_introducrion_2001, borger_abstract_2003}. 

The solution proposed in this paper aims to check a high-level specification of the system behavior before design. The chosen approach to build a model is similar to \emph{state analysis} \cite{michel_ingham_engineering_2005}, in the sense that ``states'' of the system are modeled separately and used to control the system behavior. The difference being that in \cite{michel_ingham_engineering_2005} the model is more detailed, specifying a control of hardware. Ingham developed this approach in response to several issues, similar to those encountered in BT:

\begin{itemize}
\item Subsystem-level functional decomposition fails to express the whole system behavior
\item There is a gap between the requirements and their implementation
\item The system behavior is not explicitly specified
\end{itemize}

\subsection{Method}

In the solution, the system is modeled while separating its description from its behavior. The description of the system relates here to its structure and properties, and more generally what is known regrading the system at a point in time. The system description is modeled using \emph{states}, while the behavior is modeled using \emph{modes}. The concepts of \emph{states} and \emph{modes} used here come from a previous work \cite{baduel_definition_2018}. The method developed is divided into three parts:

\begin{itemize}
\item Modeling method
\item Verification method
\item Creation of the execution model
\end{itemize}

\subsection{Case study}

The method has been experimented on a real project data, using high level specifications of a \emph{MOVIA} metro. The focus was on the train main functions when operating under normal conditions, excluding maintenance, emergencies, restricted and degraded operations and secondary use cases (e.g., comfort features) to concentrate on the train activation and driving operations. This resulted in the selection of 54 use cases divided into 12 scopes. Following the presented method, data was captured or specify in order to build an integrated model of the system and check its behavior. %The data are is presented here.

\section{Behavior description through states}

\subsection{State definition}

\emph{Types of  states}\cite{baduel_definition_2018} are modeled as finite state machines arranged in an structure of holons, similar to what is presented in \cite{dingel_modeling_2014}. A holon is en element that is both a whole, something that can exist and function independently, and a part, meaning it can be connected to other element as part of a structure. Here, each finite state machine is a holon. Rather than representing the system behavior, the holonic structure is used to established traceability between states, some being deduced from others. This allows a first form of integration by providing an evolving description of the system using correlated information.

Different types of states can be defined, each characterizing a kind of information regarding the system. For a train, we can consider:

\begin{itemize}
\item The \emph{operability}: the readiness of the train
\item The \emph{energy supply}: the source used to power the train
\item The \emph{environment}: the place where the train is operated
\item ...
\end{itemize}

Each type of state can take the value of a corresponding set of \emph{state values}. While they may appear as mere variables, those types of states are not necessarily measured or calculated, as they can express a ``known'' information, as it is the case for the operability. What truly differentiates a type of state is that its state values change depending on the target it qualifies and the adopted point of view. 

Types of state express pieces of information that have been identified and separated to characterize the conditions under which the system is used and where the different use cases can be performed. As such, the information contained in a given type of state can be abstract. Defining a type of state is valuable on its own, as the following example will show.

\begin{table}[htbp]
	\centering
	\caption{Definition of the type of state \emph{operability}}
	\begin{tabular}{|l|r|}
	\hline
	\multicolumn{2}{|c|}{\emph{Operability}} \\
	\hline
	 \textbf{Target} & The train system \\
	 \textbf{Information} & The train operability \\
	 \textbf{Context} & The train daily life-cycle \\
	 \textbf{Abstraction level} & System level \\
	 \textbf{View} & Point of view of the train \\
	\hline
	\end{tabular}
	\label{OperTypeState}
\end{table}

\begin{figure}[htbp]
\centerline{\includegraphics[scale=0.55]{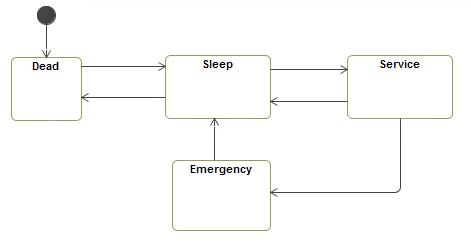}}
\caption{State values and transitions of the type of state \emph{operability}}\label{Oper}
\end{figure}

A good example of a type of state used to describe the train system is its \emph{operability}, as defined in Table~\ref{OperTypeState}. Its values and constraints on transitions are described by a state machine in Figure~\ref{Oper}. It is a type of state that indicates the capability of the train to pursue a mission. It mainly depends on the status of the train energy supply and the activation of internal systems. This information is expressed at the system level, which is treated like a black box. 

Operability is defined before making a design, and most use cases can be performed for several states of operability. It means that, on its own, operability is first an abstract type of state with no practical ways of evaluating it, is too broad to characterize use cases on its own but still provides a key information to the user regarding the train utilization and evolution. 

\subsection{States in the case study}

Fifteen types of states were established to describe the system operational condition and situation for the selected use cases. A sample of them with their possible values is shown in Table~\ref{TypesOfState}, with information regarding whether the train moves, if it is on a section with an electrical line (not neutral) and which source of energy is supplying the train systems. The \emph{State space} of each of the 15 types of states varies from 2 to 6 different values. 

\begin{table}[htbp]
	\centering
	\caption{Example of types of state describing a train operational status}
	\begin{tabular}{ccc}
	\hline
	Movement & \makecell{Neutral \\ Section} & \makecell{Electrical\\ Supply} \\
	\hline
		Moving & Neutral & Full internal supply \\
		Standstill & Not neutral & Internal supply depleted \\
		& & Line supply \\
		& & No supply \\
		& & Partial internal supply \\
		& & Shore supply 
	\end{tabular}
	\label{TypesOfState}
\end{table}

Train states characterize the train at its own level of granularity, from its own point of view, in the context of its whole life-cycle. This information can be found in part in the scenarios defined in the operability analysis, as they specify circumstances under which use cases are performed. Other information are obtained through empirical experience and requirement analysis. These information, independently from the behavior, can be constrained by physical laws, properties to be respected, inter-dependencies, etc.

\subsection{State constraints}

Without a design and working at system level, there may be an issue in measuring or evaluating some of the states. What can be done is defining possible state configurations. To do so, the different types of states can be linked by constraints and properties that condition the values (states) they can or should take in regard of each other. Those constraints are defined in relation to the system, and not to its functions. Consequently, establishing a correlation between them through constraints results in an integrated description of the system that will be navigated through its behavior.

The system behavior depends in parts on its circumstances, meaning its situation in relation to a context. They can be described by its states. The evaluation of all types of state is a configuration describing the train circumstances. 

In order to check the system behavior, it is important to define as many relevant constraints as possible on the state values forming configurations to reduce emergence. The more the system is constrained, the less unknown behaviors there will be, and the less cases there will be to consider. While over-constraining the system is a risk, it is not an issue in our setting: since checking the expected behavior is possible, any issue due to over-constraining can be detected early enough. The problem can then be solved, or in the worst case the specifications or the constraints were initially wrong or cannot be fulfilled or checked at this point. On the other hand, a lack of constraints will result in more unknown cases and will present the risk to perform analysis on irrelevant cases while overlooking errors in others.

Types of states can be used to check properties of the system. Definition of such properties can lead to the creation of more types of states, or constraints put on them. For example, a train ``visibility'' is ensured when the train has its lights systems activated under the right circumstances. It can be defined as a constraint on the other types of states to ensure that the train would be evaluated as ``visible'' when the circumstances ask for it. Such constraints come from requirements and knowledge regarding the expected system. They can constitute formal validation requirements when expressed using states.

We define two types of constraints: \emph{simple constraints} and \emph{complex constraints}. Simple constraints are defined between \emph{pairs} of states values and can be captured and specified by engineers. All possible simple constraints are considered, leading to new specifications and a first integration of the system states. Complex constraints represent known or desired constraints between three or more states values and cannot be exhaustively captured.

Let us denote $T=\{t_{1},...,t_{n}\}$ as the set of the types of states, with {n} the number of types of states defined. Every types of state corresponds to a set of possible state values: $\forall i\in \{1,...,n\},\exists k\mid t_{i}=\{s_{i1},...,s_{ik}\}$. For every state value $s$, we also denote by $s$ the logical proposition: ``the system is in state $s$''.

Two incompatibles states values $x,y$ of two different types of states $t_{i},t_{j}$ are represented by the simple constraint $\lnot(x\wedge y)$. For all types of states, we can define simple constraints as a set of clauses $simpleConsts$ such as:

\begin{equation}
simpleConsts \subseteq \{\overline{x} \vee \overline{y} | \forall i,j\in\{1,...,n\},i\neq j, \forall x,y\in t_{i}\times t_{j}\}\\
\end{equation}

The complex constraints are defined by forbidding combinations of state values taken from subsets of three or more types of states. Considering a group of types of states $t_{1},...,t_{k}$ with $k>=3$, a complex constraint $compConst$ can be defined as all combination of state values among the subsets  $t_{1}',...,t_{k}'$ such that $t'_{i} \subseteq t_{i}$ 

\begin{equation}
compConst = \wedge_{x_{1},...,x_{k} \in t_{1}',...,t_{k}'}(\overline{x}_{1}\vee...\vee \overline{x}_{k})
\end{equation}

\subsection{State constraints in the case study}

Listing the values of the different types of states, a square matrix can be created were engineers can specify simple constraints between states values. Compatible pairs of states values of two different types of states are marked by a 1 in the matrix, and by a 0 otherwise. An example from the case study is given in Table~\ref{ConstTypeStat}, using values from the types of states presented in Table~\ref{TypesOfState}. Only part of the square matrix is presented.

\begin{table}[htbp]
	\centering
	\caption{Simple constraints between types of states values}
	\begin{tabular}{c|cccccc}
		&
		\rot{Full internal supply} &
		\rot{Internal supply depleted} &
		\rot{Line supply} &
		\rot{No supply} &
		\rot{Partial internal supply} &
		\rot{Shore supply}
				\\ \hline
		 neutral & 1 & 1 & 0 & 1 & 1 & 1 \\
		 not neutral & 1 & 1 & 1 & 0 & 0 & 0 \\ \hline
		 Moving  & 1 & 0 & 1 & 0 & 0 & 0 \\
		 Standstill & 1 & 1 & 1 & 1 & 1 & 1 \\
	\end{tabular}
	\label{ConstTypeStat}
\end{table}

Complex constraints are defined as the rows of another matrix. The columns of this matrix correspond to the states values of each types of states. Each row specifies the subsets of state values involved in the constraint, represented by the indicator function (i.e, a 1 means the state is included in the subset). This list has not the ambition of being exhaustive, only expressing known properties from requirements and experience. Contrary to simple constraints, it is not practical, or even possible, to ask for engineers to think of all possible complex constraints. It also presents the risk to repeat complex constraints already induced by simple ones. Besides, specifying the simple constraints and correcting them often leads to the definition of new complex ones. Correcting a simple constraint means deleting it, as it was too strict and blocked the realization of use cases, and replacing it by a complex constraint that is more specific and carries the actual intent of the initial simple constraint. They often would not have been specified or thought of by other ways. An example of complex constraint C1 is given in Table~\ref{CompConst}.

\begin{table}[htbp]
	\centering
	\begin{tabular}{l|cccccc|cccc|cc}
		&
		\rot{Full internal supply} &
		\rot{Internal supply depleted} &
		\rot{Line supply} &
		\rot{No supply} &
		\rot{Partial internal supply} &
		\rot{Shore supply} &
		\rot{Depot} &
		\rot{Insertion line} &
		\rot{Main line} &
		\rot{Station} &
		\rot{Neutral section} &
		\rot{Not a neutral section}
				\\ \hline
		 C1 & 0 & 1 & 1 & 1 & 0 & 0 & 0 & 1 & 1 & 1 & 1 & 0 
	\end{tabular}
	\caption{Exemple of a complex constraint}
	\label{CompConst}
\end{table}

\subsection{Use case pre-conditions}

States capture the circumstances in which a use case is possible. The preconditions should capture every possible configuration in which a use case is possible, the limitations being expressed through the constraints. A precondition has a subset of authorized state values for each type of state. Considering the subsets $t_{1}',...,t_{n}'$ of the types of states $T$ for a given precondition $precond$, we have:

\begin{equation}
precond = \wedge_{i \in 1,..,n} (\vee_{s \in t'_{i}}s)
\end{equation}

\begin{table}[htbp]
	\centering
	\caption{Exemple of a use case precondition}
	\begin{tabular}{l|cc|cc|cc}
		&
		\rot{Full internal supply} &
		\rot{Partial internal supply} &
		\rot{Neutral} &
		\rot{Not neutral} &
		\rot{Moving} &
		\rot{Standstill}
				\\ \hline
		 \makecell{Wake up\\train} & 1 & 1 & 1 & 1 & 0 & 1
	\end{tabular}
	\label{ucPrec}
\end{table}

Use cases preconditions are defined in a matrix indicating which values of each type of states are compatible with their realizations. Compatible values are marked with a one, incompatible ones with a zero. Those preconditions indicate which values can and should be found in a configuration satisfying the use case preconditions, but do not imply that all combinations of compatible values are possible, as there are constraints to consider. Considering only preconditions of this form is justified by the fact that engineers can focus on the use cases preconditions one state at a time. 

\subsection{Use case pre-conditions in the case study}

An example is given in Table~\ref{ucPrec} (only authorized values have been displayed for the energy supply). Initially, waking up the train following a scenario to put it into service was not possible as a constraint indicated that a train could not be still on a neutral section. A neutral section is a section where there is no electrical supply from the line, which is the case where the train is parked. The reason for this error was that engineers made the specification while thinking of the train as performing a mission on the main line. A neutral section can be found on the main line, in which case a train should indeed not stop, but a train in a depot is also technically on a neutral section, but still needs to move on its own. This lead to the definition of types of states expressing that the train was in a mission or not, and what its environment is, as well as expressing complex constraints to enforce what was intended in the original specification. The initial specifications on their own were either incomplete or not-binding, letting developers of subsystems interpret the information and correct it.

\section{Verification method}

In order to develop an integrated model to verify and validate the system's expected behavior, it is necessary to first have proper inputs. To that end, a solution has been developed for engineers to check some predefined properties of their specifications. The solution is automatic and works like a black box: it is a script coded in R language that takes directly the matrices defined previously as inputs, without a need for other modeling activities. The technical details are presented first, the results and errors detected being discussed after.

\subsection{State constraints verification}

Two basic sanity checks are performed:

\begin{enumerate}
\item There is at least one compatible value between two state types.
\item Each state value appears in at least one possible state configuration.
\end{enumerate}

Performing the first sanity check implies checking that the simple constraint matrix is correctly filled. 

Calculating possible configuration is done by a script using applications of the \emph{graph theory} \cite{balakrishnan_textbook_2012}. The solution is intended to correspond to the industrial practice and needs, and as such is not optimal. A more elaborate solution is currently not needed considering that the calculation only take seconds. 

The script performs the following actions, logging errors at each check step:

\begin{itemize}
\item Check that the matrix is correctly filled.
\item Calculate all possible configurations according to simple constraints
\item Filter possible configurations using complex constraints.
\item Check the presence of each state value in at least one of configuration of the filtered list.
\end{itemize}

\subsection{Use case preconditions verification}

For every use case we check that:

\begin{itemize}
\item Its precondition admits at least one possible configuration regarding the state constraints
\item Each authorized state value appears in at least one possible state configuration.
\end{itemize}

The script performs the following actions, logging errors at each check step and for each use case:

\begin{itemize}
\item Calculate possible configurations
\item Check the presence of each authorized state value in a least one of the possible configurations.
\item Filter the configurations with complex constraints. Check again for the existence of a solution.
\item Check the presence of each authorized state value in a least one of configuration of the filtered list.
\end{itemize}

Simple constraints and complex constraints are applied and checked separately to facilitate their analysis and correction. 

\subsection{Results}

The results can easily be formatted and customized, in our case an Excel file. The case study showed that:

\begin{itemize}
\item Out of 15 types of states, 2 pairs initially lacked authorized values between them.
\item Out of 45 state values, 6 were not initially included in any possible configurations.
\item 5 more complex constraints were defined after correcting the simple constraints.
\item Out of 54 use cases, 7 initially lacked authorized values.
\item 13 use cases did not initially admit a single configuration as solution.
\item 51 use cases had unused values, for a total of 148 cases.
\end{itemize}

The lack of authorized values between types of states or in preconditions were simple omissions. State values not included in any possible configurations were due to the following errors:

\begin{itemize}
\item State were ill-interpreted by the engineers.
\item Engineers adopted a point of view that was too narrow, overlooking  specific cases were some states values were compatible.
\end{itemize}

Nearly all use cases had unused state values, meaning state values authorized in the preconditions but not present in any of the related possible configurations. In order of increasing severity, it could mean that:

\begin{itemize}
\item a given state was deemed possible in the preconditions but was not,
\item the use case should have admitted a configuration with this state but its preconditions were too narrow,
\item there was an issue in the way the constraints were defined, blocking possible configurations. 
\end{itemize}

As the states are used for the preconditions of all use cases and their constraints are used in the calculations of all possible state configurations, an error in their definition is where it has the most severe impact.

The method proved that when integrating specified information on current validated steps of a project, there were in fact many errors and misunderstanding that would have to be corrected later on. Those errors were detected here at an earlier stage in the process. In addition, this analysis provides new or proper specifications as opposed to partial or informal ones.

\section{Execution model}

According to our approach, the fundamental unit for organizing the system description is the state, and the fundamental unit for organizing the behavior is the mode. The behavior is modeled by hierarchical state machines, here SysML statecharts, where ``state'' modeling elements correspond to our concept of mode. Each mode can be activated after checking that the system state configuration allows it and that the right sequence of activities has been executed. A mode here characterizes use cases of the system while specific conditions on the system state are true. 

\subsection{Holonic structure for states}

The system description is modeled following this process:

\begin{enumerate}
\item Create a SysML block for each type of states.
\item Model each type of state as a non-hierarchical statechart inside each block.
\item Create a signal for every transition of state values.
\item Create the structure around state blocks by creating and linking ports
\item Enclose all state in one SysML block contained by the system block.
\item Define inputs ports for signals updating state values from environment or the system behavior.
\end{enumerate}

\subsection{Structure of the behavior}

There is a need for both a specification and an executable model of the whole system and its behavior. In order to integrate the specification, it should be possible to specify dynamic aspects of the use cases. It requires knowing which use case can be realized in a given situation and how their possible realization evolves. Conditions enabling the realization of a use case correspond to our definition of a mode. The preconditions defined earlier enable to know when a use can can be performed, and can be considered as a basis for defining modes. We will now define a structure of modes to analyze, integrate and model the behavior. We define several types of modes:

\begin{itemize}
\item Use case mode: conditions the realization of a given use case
\item Scope mode: a mode defined by a precondition composed of subsets that are the union of authorized values in all preconditions of all use cases modes under the corresponding scope. 
\item Abstract mode: conditions the activation of one or several scope mode.
\end{itemize}

The way the scope modes and abstract modes are defined is potentially larger than the disjunction of use case preconditions and scope mode conditions they refer to, in order to cover a broader context and follow the evolutions of use case transition. This is also a way of ensuring that we have implication relationships between the different modes, something we need to build a structure around them.

Use case modes directly condition the execution of use cases. Other types of modes only condition them in an indirect way by conditioning the use case modes or the modes containing those. A mode is defined by the use cases it characterizes (directly or not) and the conditions in which it is activated. As the relationship between modes and use cases is established, the main characteristic needed for the definition of new modes is the conditions for which they are activated.

The conditions for a mode can be considered as a group of subsets of each types of states's values. As long as all state values of a state configuration are part of these subsets, the mode is active. Given $T=\{t_{1},...,t_{n}\}$ the set of all types of states, a mode has the same structure as a UC precondition (see section 2.5) and is defined by a set of subsets $t'_i \subseteq t_i$.

The behavior can be modeled using statecharts. As the use cases are managed in scopes allocated to different engineers and that they are too numerous to be put in one statechart, all use cases modes should be put under global modes corresponding to their scope, where they are to be modeled in a corresponding statechart. Considering  $\{t_{1}',...,t_{n}'\}$  the conditions of a scope mode, we define the conditions of the k use cases under this scope as $\forall j\in\{1,...,k\}, \{t_{j1}',...,t_{jn}'\}$. We have:

\begin{equation}
\forall i \in \{1,...,n\}, t_{i}'= \cup_{j=1}^{k}t_{ji}'
\end{equation}

In order to integrate the statecharts defined in each scope, there needs to be a way to evaluate whether the different scopes modes are activated or not. We propose to create a structure of implications enabling to determine activated modes by evaluating their conditions. 

Satisfying the conditions of a use case mode means the conditions of its scope mode are satisfied: activating a use case mode implies activating its scope mode. In the same way, some scope mode could imply others, which is the basis for our implication structure. Some scope modes could also have the same conditions, in which case we create one statechart in each scope to specify the behavior but only define one corresponding scope mode.  

All scopes modes may not be linked by implication relationships, in which case we define abstract modes. Abstract modes are obtained by the union of two modes preconditions. We only define abstract modes for pairs of scope modes that do not imply any other.

\subsection{Structure of modes in the case study}

\begin{figure}[htbp]
\centerline{\includegraphics[scale=0.6]{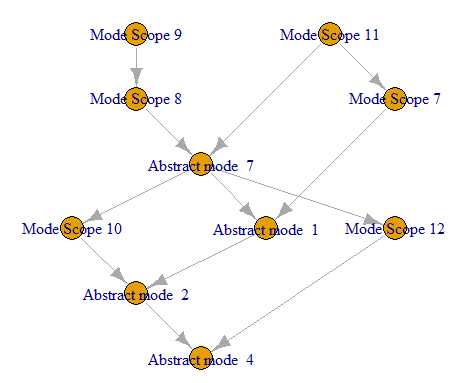}}
\caption{Implication structure of scope modes and abstract modes}\label{modStr}
\end{figure}

The structure of modes is generated thanks to a script. The result of its application on the case study is shown in Figure~\ref{modStr}, using the modes of 6 scopes for visibility. As the modes imply each other, a path of implication correspond to preconditions that are more and more specific. Keeping only the longest paths, obtained by transitive reduction (which as been applied in the example), correspond to progressive definitions of increasing details in the preconditions that are each evaluated once.

\section{Method}

The system is modeled while considering separately its states and its behavior. 
The same approach is then applied on its elements, detailing the behavior while maintaining states and their traceability. The goal is to specify and check an integrated system and its behavior that would otherwise be either unspecified or emergent depending on the level of development. The method developed follows this process:

\begin{enumerate}
\item Define types of states providing information on the system at its own level of granularity.
\item Define simple and complex constraints between the values of different types of states to correlate the information.
\item Define enabling circumstances of the system use cases using the system states.
\item Check all states constraints and use cases preconditions.
\item Generate the structure of modes.
\item Build the system description and behavior models.
\end{enumerate}

Once an integrated model of the system is obtained, more V\&V activities can be performed, such as simulation. 
The model evolves by completing its elements models, their description (states), and the behavior detailed amongst all elements.

The information expressed by the states can evolve internally through deduction amongst types of states.
They also can be modified by the behavior or be communicated by the environment. 
The use cases, however, express interaction with the system processed through its behavior, and do not directly change the system state. 
This way, both the behavior and the actions on the system can evolve and be detailed without modifying the description of the system. 

\section{Conclusion and perspectives}

Our results helped proving that our definition of system states enables to specify the system and its behavior, its evolution and the conditions under which use cases can be performed. 
The most direct benefit of this work is a means to check the preconditions of the use cases, not just individually but as part of a constrained whole. 
The simplicity of the solution makes it available to any engineer working on use case specification and can be easily integrated in an existing method, process or tool. 
One of the gains expressed by the BT engineers who experimented our approach was that preconditions were clearer but more importantly centralized, and instead of repeating similar information and preconditions in many requirements, those information are expressed by the same elements, the states.

The method presented does not guarantee to cover all states of the system, nor all details of its behavior. It simply aims at specifying the integration of the system to check its coherence and reduce emergence in its properties and behavior. The method remains to be deployed on a whole project for validation purposes. It also only covers discrete event phenomena. 

The next step is to perform V\&V activities on the model and then conduct the same analysis on a developed design using part of the model to check the continuity.

\section*{Acknowledgment}

This work is supported by Bombardier Transport SAS and the ANRT CIFRE grant \#2016/0262.

\bibliographystyle{splncs04}
\bibliography{ArxivPaperRB01bib}

\end{document}